\def\year{2021}\relax
\documentclass[letterpaper]{article} 
\usepackage{aaai21}  
\usepackage{times}  
\usepackage{helvet} 
\usepackage{courier}  
\usepackage[hyphens]{url}  
\usepackage{graphicx} 
\usepackage{booktabs}
\usepackage{natbib}  
\usepackage{caption} 
\title{Local News Online and COVID in the U.S.: Relationships among Coverage, Cases, Deaths, and Audience}
\author {
    Kenneth Joseph,\textsuperscript{\rm 1}
    Benjamin D. Horne,\textsuperscript{\rm 2}
    Jon Green,\textsuperscript{\rm 3}
    John P. Wihbey \textsuperscript{\rm 4}  \\
}
\affiliations {
    \textsuperscript{\rm 1}Computer Science and Engineering, University at Buffalo, Buffalo, NY, USA\\
    \textsuperscript{\rm 2}  School of Information Sciences,  University Tennessee Knoxville, Knoxville, TN, USA\\
    \textsuperscript{\rm 3} Network Science Institute, Northeastern University, Boston, MA, USA\\ 
    \textsuperscript{\rm 4} School of Journalism and Media Innovation, Northeastern University, Boston, MA, USA\\    
kjoseph@buffalo.edu, bhorne6@utk.edu, jo.green@northeastern.edu, j.wihbey@northeastern.edu
}
\begin{document}

\maketitle

\begin{abstract}
 We present analyses from a real-time information monitoring system of online local news in the U.S. We study relationships among online local news coverage of COVID, cases and deaths in an area, and properties of local news outlets and their audiences. Our analysis relies on a unique dataset of the online content of over 300 local news outlets, encompassing over 750,000 articles over a period of 10 months spanning April 2020 to February 2021. We find that the rate of COVID coverage over time by local news outlets was primarily associated with death rates at the national level, but that this effect dissipated over the course of the pandemic as news about COVID was steadily displaced by sociopolitical events, like the 2020 U.S. elections. We also find that both the volume and content of COVID coverage differed depending on local politics, and outlet audience size, as well as evidence that more vulnerable populations received less pandemic-related news.
\end{abstract}

\section{Introduction}
\noindent 
As health and medical professionals continue to fight COVID on the front lines, a different battle is playing out on Americans' screens. Public health officials and politicians, aiming to spread critical health-related information and to shape the narrative around the pandemic, are fighting for space and attention in both new and traditional media. In order to understand fulfilled and unfulfilled information needs during the crisis, tools and data are needed to rapidly characterize what information the public is receiving, and how that information varies across populations. 

To this end, scholars have begun to analyze the information environment surrounding COVID, showing partisan variation in coverage in national news outlets in the U.S. \cite{muddimanCableNightlyNetwork2020,krupenkinIfTreeFalls2020}, and how urban/rural divides in news have been associated with behavior \cite{kimEffectBigcityNews2020} during COVID.
However, despite the critical role they have played in information consumption during COVID \cite{ritterPandemicNewsAttention2020,kimEffectBigcityNews2020} and their consistent track record of shaping opinion and behavior \cite{schulhofer-wohlNewspapersMatterShortRun2013,hayesLocalNewsGoes2015,gentzkowEffectNewspaperEntry2011},  no work has yet looked at what information has been produced from local news outlets during the pandemic, or how that coverage has varied across outlets.  Local news, moreover, is important because at least early in the pandemic, the majority of Americans turned to local news outlets to understand the impact of COVID in their community \cite{frankCOVID19DrivesTraffic2020}.  
Further, almost no published work has looked at content of any kind during later stages of the pandemic, especially during the deadly "third" wave that began in November 2020. How the media responded to the prolonged impact of the pandemic is therefore still largely unclear.

We address these questions using a unique dataset of the complete set of 757,053 news articles written by 310 local newspaper outlets in the U.S. who provide RSS feeds of their online content. Our dataset spans a breadth of local communities around the country (see Figure~\ref{fig:fig1b}), and captures articles written between April, 1st, 2020 and February, 17th, 2021, allowing for analyses regarding local information environments over a long time span. To the best of our knowledge, our data is the most comprehensive corpus of local news outlets available for research. While there are similarly complete news datasets at a national level \cite{norregaard2019nela,gruppi2021nela}, no such collections exist at the local level.

Using these data, we address two primary research questions: 
\begin{itemize}
    \item \textbf{RQ1}: What factors are associated with the degree of local news coverage, over time, of the COVID-19 pandemic?
    \item \textbf{RQ2}: What were the primary COVID-related \emph{topics} covered during the pandemic, and how did this coverage vary across outlets?
\end{itemize}

\begin{figure}[t]
    \includegraphics[width=.45\textwidth]{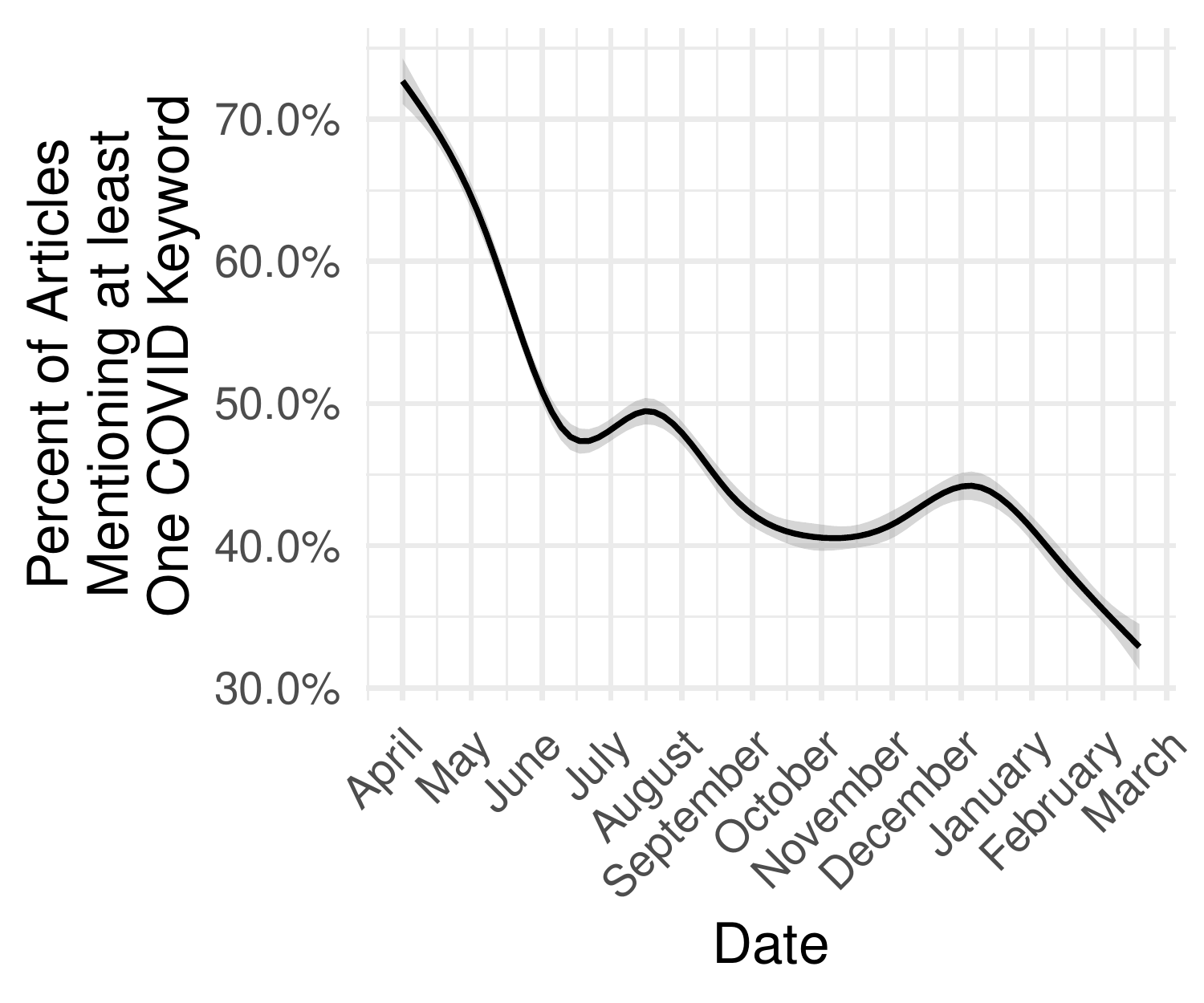} \\
    \caption{The average percentage of articles about COVID over time for all 310 news outlets, fit using a generalized additive model.}
    \label{fig:fig1a}
\end{figure}

\begin{figure}[t]
    \includegraphics[width=.49\textwidth]{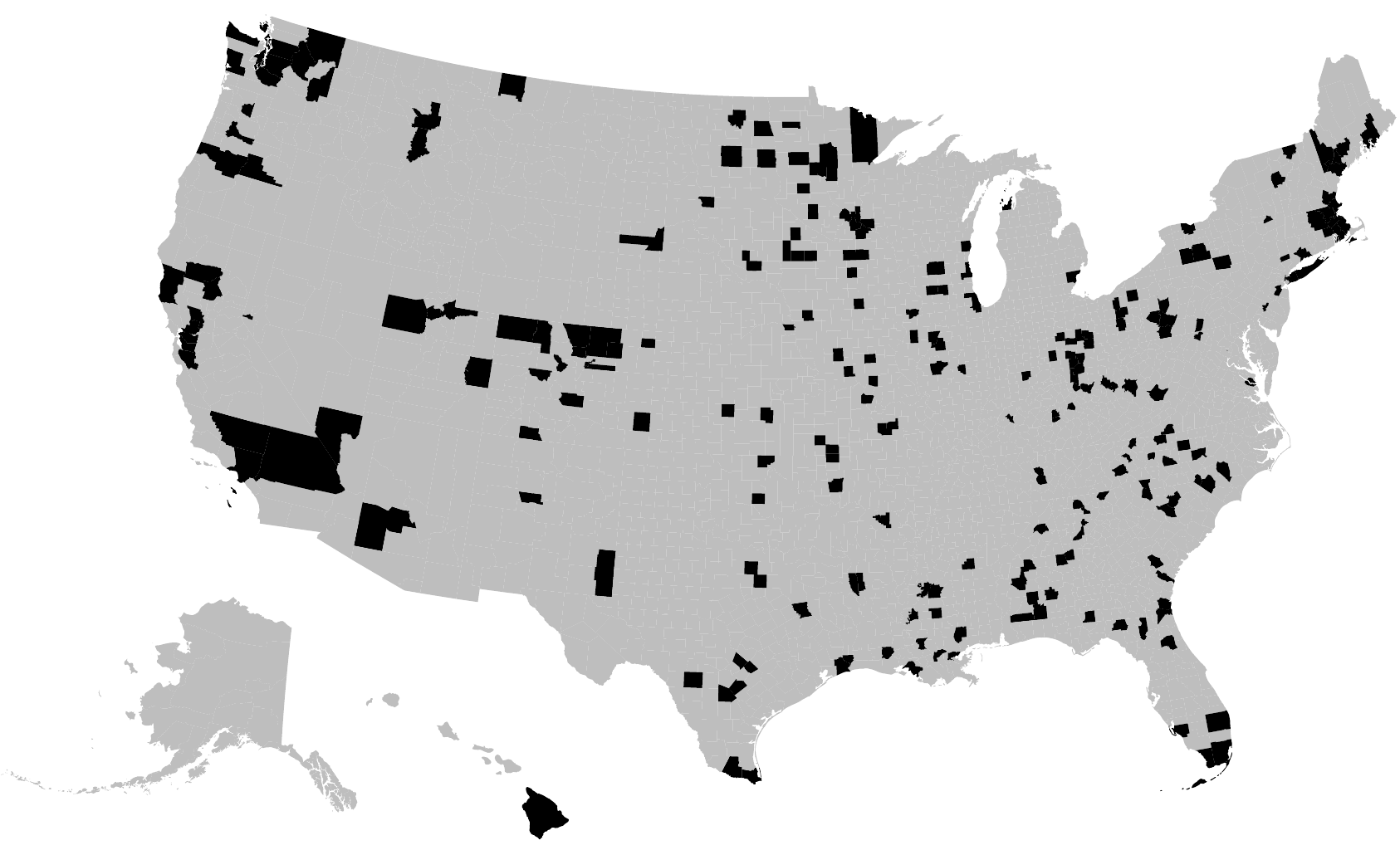} \\
    \caption{A map of the spatial coverage of our dataset. Counties colored black contained at least one local news outlet that we studied. }
    \label{fig:fig1b}
\end{figure}

With respect to \textbf{RQ1}, it would be reasonable to expect that coverage of COVID varied with the severity of the pandemic in the local area. However, many other factors \cite{mccolloughPortraitOnlineLocal2017}, e.g. economic issues \cite{petersonPaperCutsHow2021}, are at play in determining what local news outlets choose to cover. For example, Figure~\ref{fig:fig1a} shows that coverage of COVID was not equally distributed over time---on average, local newspaper agencies shifted their attention away from COVID as time progressed. We use a set of COVID-related keywords informed by prior work \cite{shugars2021pandemics} to identify articles that were or were not related to the pandemic. Using the county an outlet is located in as a proxy for the outlet's audience, we then estimate the relationship between the percentage of articles written about COVID each week by a given outlet and A) the number of local and national COVID cases and deaths during that week, B) outlet popularity, and C) the politics of and level of COVID-mortality risk in the outlet's audience.

However, studying overall coverage of the pandemic does not tell us what facets of the pandemic were covered, nor what non-COVID related content displaced COVID coverage over time. We address this question with \textbf{RQ2}. Namely, we use structural topic modeling \cite{roberts2014structural} to understand the broad COVID and non-COVID related themes covered over time in our dataset. We then connect the degree of topic-specific COVID coverage to outlet and audience-related factors to better understand how, even within coverage of the pandemic, topical focus varied.

Our work presents the first detailed look at what was covered by local news outlets during the pandemic, and how that coverage varied based on the outlet’s audience and the situation on the ground. We also provide tools and data that can be used to continue to explore, in near real-time, how narratives about COVID are taking shape in different communities across the U.S. As such, while our analysis focuses only on the United States, we provide a generalizable methodology that can be applied to data from other areas of the world. To this end, complete replication materials are publicly available\footnote{\url{https://github.com/kennyjoseph/covid_localnews_public}}.

\section{Related Work}

Our paper draws heavily on two existing bodies of literature; a long-standing literature on local news in the United States, and a rapidly growing literature on the information environment that has grown around COVID-19. We address each of these separately below.

\subsection{Local News in the U.S.}

Local news has increasingly been defined as \emph{news which primarily serves communities at the county level, or a subset of that geographic area}. For a sense of scale, there are 3,143 counties in the United States. A central focus of the media research community studying local news has become the problem of “news deserts” – geographical units that lack consistent coverage from news institutions; foundational work continues to use the county-level as a fundamental unit of analysis \cite{abernathy2018expanding}. We use this same model for analyzing local news in the present work.
 
The relationship between local news coverage and its responsiveness to various factors, such as national events, has been the topic of substantial research. Recent analysis of local television coverage, for example, has noted an increase in the influence of the national news agenda, with local television mirroring more national news items and featuring fewer genuinely local stories, and a general right-ward ideological slant \cite{martinLocalNewsNational2019}. It is worth noting that the nationalization of elections and politics generally is an increasing trend observable in voter behavior \cite{meluskyWhenLocalNational2020}.

These patterns of nationalization have arisen hand-in-hand with a rapid decline in local news. As local news sites continue to close at an alarming rate \cite{darrLocalNewsCoverage2021}, new evidence suggests that stronger local news environments may attenuate these national trends such as political polarization \cite{moskowitzLocalNewsInformation2021, darrHomeStyle2021}. Likewise, research has shown that the partisan slant and nature of media ownership in local news can affect citizen views of national officeholders and candidates \cite{levenduskyHowDoesLocal2021}. Further, there is worrisome evidence that the decline in local news coverage observed in many communities is diminishing citizen engagement in civic issues \cite{hayesDeclineLocalNews2018}.

Nonetheless, many Americans still \emph{do} pay attention to local news. Sixty-four percent of Americans in 2018 said they got some news from their local paper \cite{pewresearchcenterLocalNewsAmericans2019}, much of this from online content. This number jumped significantly during the first months of COVID \cite{frankCOVID19DrivesTraffic2020}. Qualitative research has suggested that local media audiences can be understood as having a range of motivations for consumption: from an “aspirational” impulse of self-improvement to a “utilitarian” one of improving specific decisions \cite{nwlocalnews}. As local news has declined across the United States, more research has begun to examine the underlying business model and study concepts such as “willingness to pay,” or the structure of incentives that prompt persons to consume local news, or to forgo it. A chief area of concern is a perception that local news is an inferior good \cite{kim2021search}.

All of this suggests that local news is a vital, if struggling, part of the larger media ecosystem. In terms of the relationship between public health issues specifically and local news coverage, there is a large body of research with a wide variety of foci. In surveying this literature, scholars have noted that local media serve multiple functions in this regard, including that of: surveillance (information provision); interpretation (sense-making around events and their context and meaning); socialization (cultivating norms); and maintaining public attention \cite{gollustTelevisionNewsCoverage2019}. However, there is evidence that where national policy issues filter down to the local level, news media often focus on political controversy, rather than substantive information that could help citizens make decisions \cite{gollustLocalTelevisionNews2017}. Further, some researchers have found that coverage can reflect racial bias, with less news attention being directed to diseases that disproportionately affect communities of color \cite{armstrongWhoseDeathsMatter2006}. There are also ongoing concerns that the decline of local news may affect the ability of public health experts, who often depend on news reports, to perform robust surveillance \cite{branswellWhenTownsLose2018}.

\subsection{The Information Environment around COVID-19}

As noted above, a large and rapidly growing literature is building around the ways in which information was obtained about COVID-19. Much of the work examining the production and consumption of information regarding COVID-19 centered on how such information has spread online, both in general \cite{shugars2021pandemics} and, in many cases, specifically with respect to misinformation \cite[e.g.][]{zarocostas2020fight, yangInfodemic2021}. Others work examines how more traditional forms of media have shaped the narrative around the pandemic, such as national cable news
\cite{muddimanCableNightlyNetwork2020,krupenkinIfTreeFalls2020} and more local news \cite{kimEffectBigcityNews2020} -- as well as news interest in general \cite{ jurkowitzPewPartisanInterest2021}. 

The present work addresses a unique middle ground, studying how online articles from local news sites---which play a role in the online information environment but through a traditional media source---provided information about the pandemic. Initial examination of how the availability of local news may have influenced health behaviors during the early stages of the pandemic produced mixed results. As communities began implementing restrictions on economic and social activity, interest in local news websites increased \cite{koezeLocalNewsInterestCOVID2020, bhargavaCoverageVsCases}. In addition, local news availability was associated with changes in mobility patterns \cite{fischer2020LocalNewsAvailability}. However, these associations could be accounted for by additional factors such as public policies and Democratic vote share in 2016 \cite{fischer2020LocalNewsAvailability}, suggesting that national trends were overwhelming any direct influence that local news might have otherwise had at the time. Here, we examine the \textit{content} of online local news, and extend the time span of analysis beyond the early stages of the pandemic -- both of which could speak to why this was the case.

\section{Data and Methods}

\begin{figure}[t]
    \centering
    \includegraphics[width=.48\textwidth]{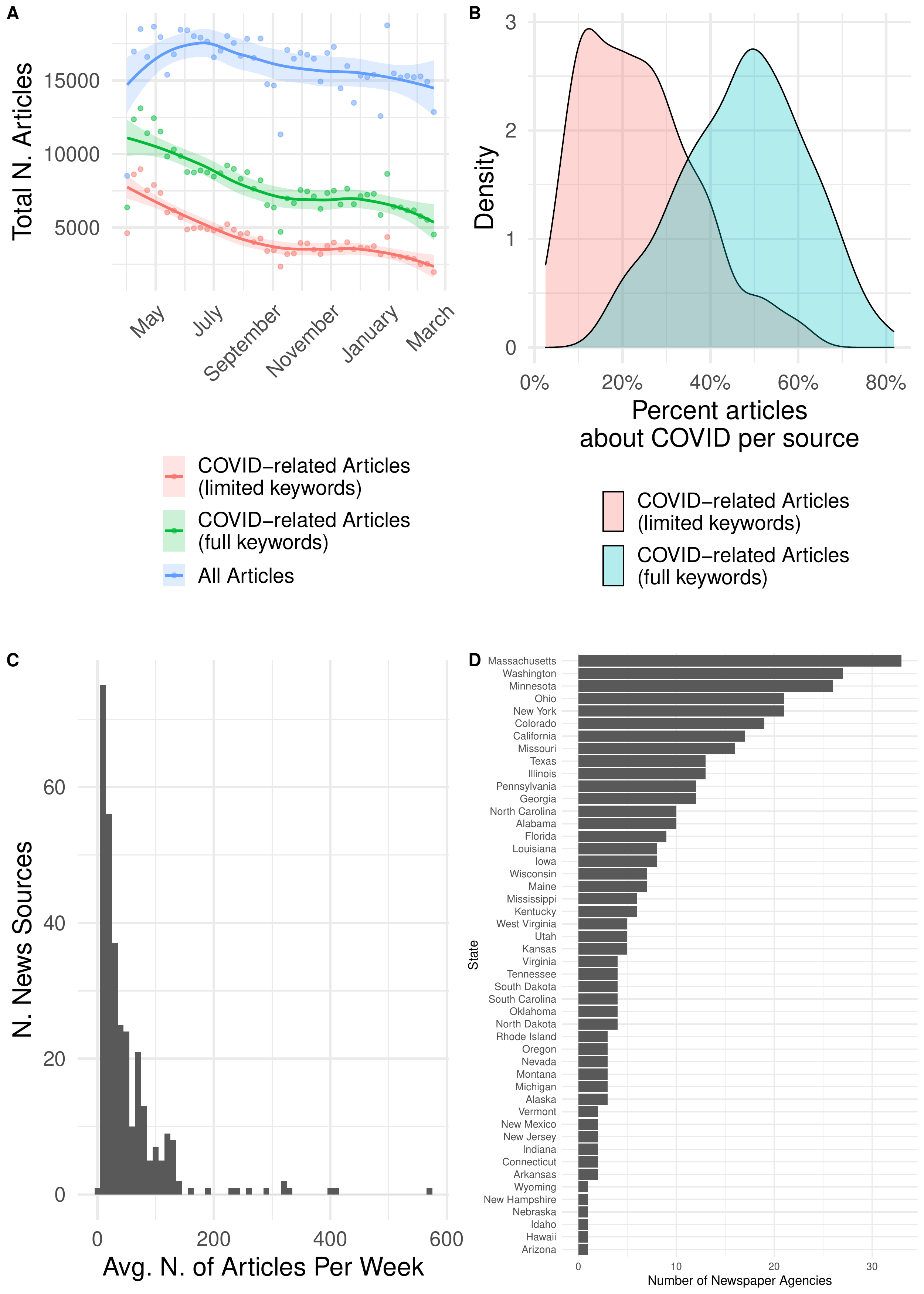}
    \caption{Summary statistics of our local newspaper dataset; figures are described in more detail in the text.}
    \label{fig:perc_covid}
\end{figure}

To address \textbf{RQ1}, regarding the link between overall COVID coverage, case/death rates for COVID, and properties associated with outlet audience, we take four steps. First, for each article in our corpus of local news articles, we use a keyword filter to determine whether or not the article discusses COVID. Second, we associate each outlet with the county in which it is headquartered, using this as a proxy for local conditions and audience. Third, we construct weekly case and death rates for COVID, and construct a number of proxy variables for outlet audience. Finally, using standard regression approaches, we look at associations between weekly COVID coverage, case/death rates, and audience proxy variables.

To address \textbf{RQ2}, we first run a structural topic model \cite{roberts2014structural} on our full news corpus. We then manually identify COVID- and non-COVID related topics, and look at how they trended over time. Finally, we associate topic prevalence at the outlet level with the same audience proxy variables used above. The rest of this section describes these data and methods in more detail.

\subsection{Local News Data}

The main data source for our paper is a large corpora of articles scraped from the RSS feeds of local news websites. In order to have a geographically diverse dataset, we begin with a publicly available list of approximately 3,300 local news outlets in the United States.\footnote{This list, at \url{www.50states.com/news/}, is one of several publicly available social studies classroom resources on the site, which has been online since 1996.} This initial list of local news outlets is then filtered down in two ways. First, we remove the 2,732 outlets in this list whose websites were now defunct, or who did not have an RSS feed. Second, we removed an additional 12 websites that were not local news. This step was completed by a paper author who is an expert in local news. Sites were removed that did not match the definition of local news provided above---namely, that it was a) a news site, b) primarily targeting a local audience and c) that the local audience was primarily described geographically. We therefore removed sites that were 1) oriented specifically towards university students, 2) explicitly national outlets, or 3) were not news-oriented. Importantly, these criterion are somewhat subjective, and the list we use may not include all local news outlets; we discuss in the limitations section of this work potential issues that could therefore have arisen in our sampling procedure.

After these two filtering steps, we then crawled the RSS feeds of the remaining 556 local news outlets in the United States twice daily starting on April 4th, 2020 until February 17th, 2021. As the RSS feeds are crawled, we follow the URLs to the web pages with full article text, and scrape that text. Importantly, we were not collecting the snippets of articles that many RSS feeds display, but rather the full text that the URls point to. 

After the data was collected, we constructed a subset of local news outlets to analyze for the present work. First, we considered only outlets that produced more than 50 articles in this time frame. Second, a small number of outlets were filtered out because we 1) could not obtain COVID case count data for the relevant county (see below), 2) could not obtain data on 2016 vote share (see below), 3) published primarily in Spanish or 4) had less than 10\% or greater than 95\% COVID-related articles for the source. The reason for the latter decision is because certain sites decided to at some point create a separate RSS feed for COVID content, which we either ended up crawling (in which case COVID-related articles were over-represented) or not crawling (in which case COVID-related articles were under-represented), depending on if this separation occurred at the start of our data collection or after. Articles from these sites were manually analyzed before removal to ensure this was the issue, as no clear pattern emerged in what kinds of local news outlets decided to split their RSS feed in this manner. The application of these filters left 310 outlets to analyze. 

The full list of 310 outlets we use in our analysis is publicly and anonymously available at: \textit{LINK REMOVED FOR BLIND REVIEW}. Here, we provide a number of useful summary statistics. Figure~\ref{fig:perc_covid}A) shows the total number of articles over time across all outlets (blue), as well as the subset of those that were identified as being COVID-related by the full filter (green) or limited filter (red) (see the section immediately below for definitions of these filters). Values are fit using a generalized additive model to show the trend.  Figure~\ref{fig:perc_covid}B) shows the distribution across outlets of the overall percentage of articles the outlet devoted to COVID coverage. Results are again shown for both the full and limited keyword filters. Figure~\ref{fig:perc_covid}C) shows the distribution of the average number of articles per week across the outlets studied. Finally, Figure~\ref{fig:perc_covid}D) shows the distribution across U.S. states in the number of outlets.


To the best of our knowledge, our data is the largest existing, publicly available collection of local news content. The closest publicly available dataset to ours is from MediaCloud \cite{roberts2021media}. While our dataset shares outlets in common with MediaCloud, the local U.S. outlets covered by the latter are embedded almost exclusively in large population centers, while our dataset covers local media outlets from both large and small population areas.

\subsection{Identification of COVID-related News Articles}
We use two approaches, both keyword-based, to identify COVID-related articles: a {\it full filter} approach and a {\it limited filter} approach. The full filter approach begins with a set keywords from prior work on COVID and Twitter \cite{gallagher2020sustained,shugars2021pandemics}. From this list, we remove keywords that are specific to Twitter; namely hashtags and \@-mentions, and retain all others. \citet{shugars2021pandemics} show that this keyword list is effective in capturing COVID-19 related Twitter content; their manual evaluation on a random sample of tweets showed over 90\% accuracy when using this keyword list to classify tweets as COVID-related or not.\footnote{The list used in the present work is available here: \textit{LINK REMOVED FOR BLIND REVIEW}}. Still, it is possible that this keyword list does not translate well from social media to news articles. Therefore, as a robustness check, our limited filter tags articles as being COVID-related only when they use one of the following terms: covid, covid19, coronavirus, sars-cov-2. We present results for both filters, showing that main findings are robust to the use of either one of these keyword lists.


\subsection{Associating Outlets with Audience Proxies}

In order to connect patterns in the publication of news to the composition of the outlet's audience, we linked each local news outlet to the county in which its headquarters is situated in.  We then link each outlet in our dataset to four different county-level datasets. 

First, we used data from \emph{The New York Times} on COVID case counts and numbers of deaths.\footnote{\url{https://github.com/nytimes/covid-19-data}} Note that these data do not separate out different counties in the New York City area, and thus we assign case and death counts for all outlets in this area to the single New York City counts in the New York Times dataset.  We also aggregate these values up to get state and country-level case and death counts. Because counts are of course impacted by population size, we follow \citet{warshawFatalitiesCOVID19Are2020} an translate these counts to (logged) rates of cases and deaths per 1,000 citizens. To do so, we use 2019 population estimates of U.S. counties from the Census Bureau.\footnote{\url{https://www2.census.gov/programs-surveys/popest/datasets/2010-2019/counties/totals/}}

Second, we consider the political leanings of the counties in which each outlet is situated. To do so, we link each outlet to the 2020 vote share for Trump vs. Biden in the county the outlet was situated in.\footnote{ \url{https://raw.githubusercontent.com/kjhealy/us_elections_2020_csv/master/results_current.csv}}. As a robustness check, we also link to 2016 vote counts for Trump and Clinton.\footnote{\url{https://github.com/MEDSL/2018-elections-unoffical/blob/master/election-context-2018.md}}. The correlation between these two values at the county level is .987; we therefore focus only on 2020 returns. However, the latter dataset does not include values from Alaska, and hence we remove Alaska-based local news outlets from our analysis. 

Third, we move beyond case and death counts to also consider the \emph{risk} of the local news outlet's audience to COVID-19. In this sense, we consider both the actual and potential impact of COVID-19 on the counties in which local news outlets were situated. To quantify this risk, we used the recently released \emph{Community Resilience Estimates} constructed by the U.S. Census Bureau. These Community Resilience Estimates were constructed using 2018 data from the American Community Survey, National Center for Health Statistics National Health Interview Survey, and Population and Housing Unit Estimates. Their intent is to ``measure the ability of a population to absorb, endure and recover from the impacts of disasters... such as COVID-19''  \cite{uscensusbureauCensusBureauReleases2021}. The Census Bureau provides estimates of the percentage of individuals in each county that had 0 risk factors, 1-2 risk factors, or 3+ risk factors for COVID. The data and its full description are publicly available\footnote{\url{https://www.census.gov/data/experimental-data-products/community-resilience-estimates.html}}.

Finally, in addition to these county-level variables, we link each local news outlet to its ranking in Amazon Alexa's database of website popularity, accessible through a free API\footnote{\url{https://aws.amazon.com/marketplace/pp/Amazon-Web-Services-Alexa-Top-Sites/B07QK2XWNV}}. Three outlets, comprising less than 1\% of our dataset, were not found in the Alexa database. We imputed the minimum rank of all other outlets for these three cases.

\begin{figure}[t]
    \centering
       \includegraphics[width=.48\textwidth]{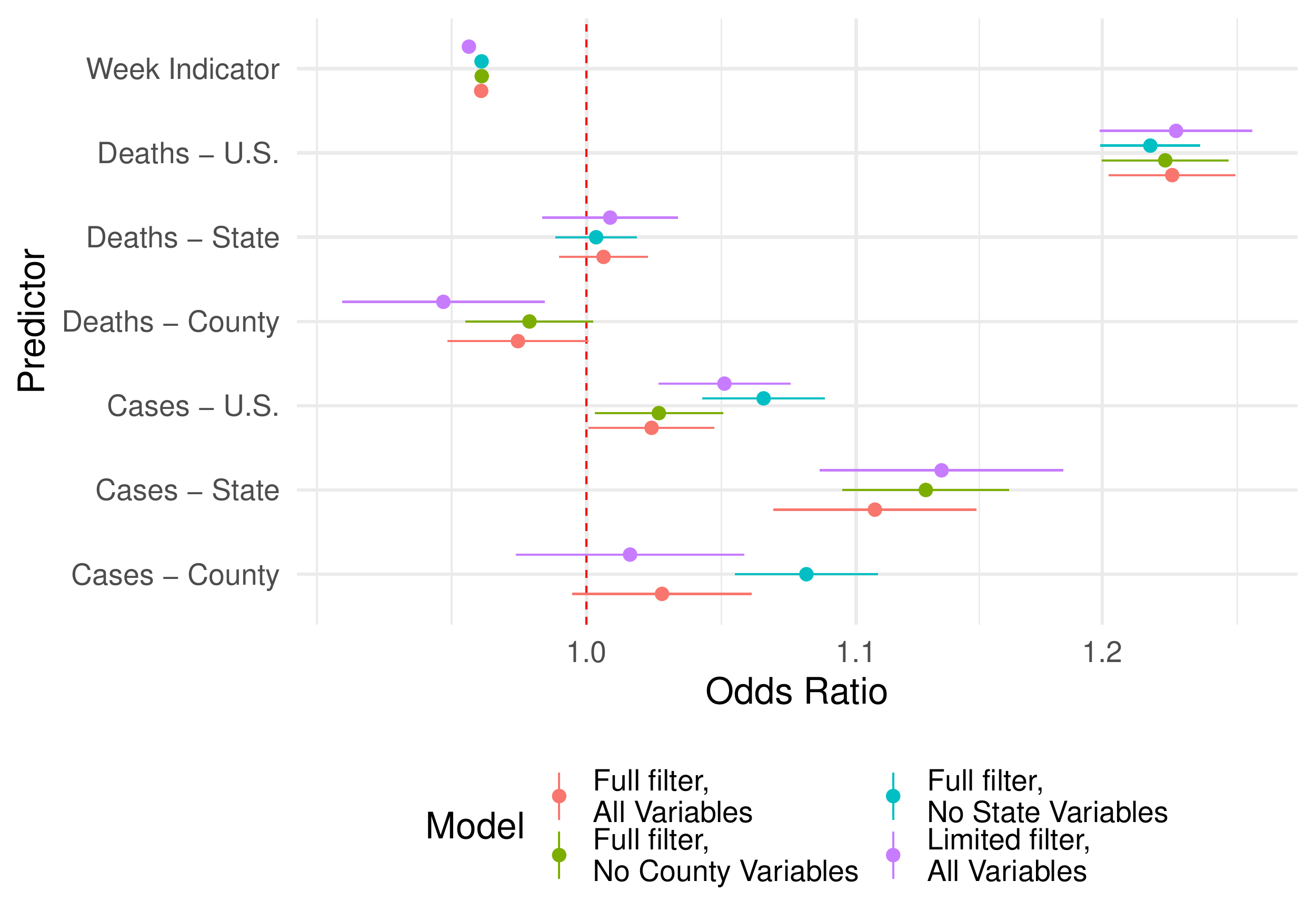} 
    \caption{Change in the odds (y-axis) of an article being about COVID-19 in a given week for a given news outlet with a 1 standard deviation change in a given predictor (x-axis). Error bars are 95\% standard errors with outlet-level clustering accounted for.  Four models are considered - two with all relevant independent variables, one without county-level case and death rates, and one without state-level case and death rates, because county and state level rates are correlated in our data.
    }
    \label{fig:rq1a}
\end{figure}

\begin{figure}[t]
    \includegraphics[width=.48\textwidth]{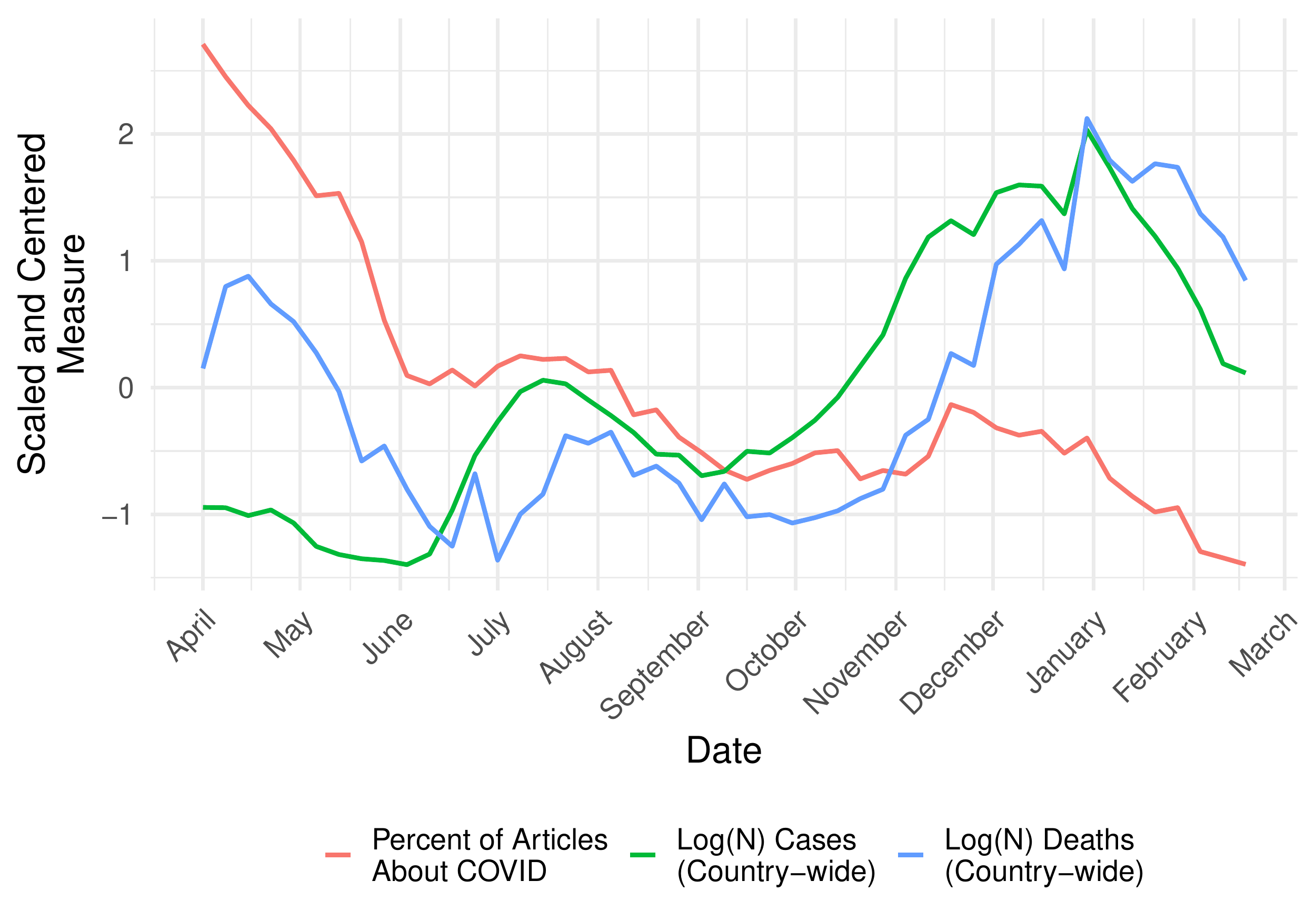} \\
    \caption{Changes in percent of COVID articles over time across all outlets (red line), and the log number of COVID cases (green) and deaths (blue) over time. All measures are scaled and centered so they align on a single axis.}
    \label{fig:rq1b}
\end{figure}

\subsection{Identifying topics in local news coverage}
Finally, in order to understand if there were differences in how local newspapers covered COVID, we constructed a Structural Topic Model (STM) \cite{roberts2014structural} on all news articles in our data set, with time as a covariate in the model. The STM is part of a larger family of \emph{topic models} used to rapidly extract themes from text data; for a useful primer on topic modeling, see \citet{blei2012probabilistic}.  Briefly, a Structural Topic Model (STM) is a generative model of word counts that allows for the use of document-level metadata. Hence, rather than only generating a topic-word distribution per document, we can also associate those topic-word distributions with other variables (in this case, time) which can be used to help guide the model to topics that are more coherent. The STM is particularly useful, relative to other topic models, when covariates exist that are a priori known to have an effect on what themes will exist in different documents. In the analysis of news data, a common such covariate is time, used here. Specifically, we treat time as a continuous variable estimated with a b-spline basis (e.g. a variable that can have a non-linear association with the distribution of how prevalent each topic is over time). 

Notably, we select the structural topic model as opposed to more recent topic modeling frameworks that incorporate more recent innovations in the natural language processing literature, in particular those that incorporate new and more rapid approaches for model optimization \cite{card2018neural} and those that incorporate contextualized word embeddings \cite{bianchi-etal-2021-pre}. We opt for the STM over these other approaches for two reasons. First, the STM has been widely applied and validated for use in the social sciences, relative to these newer models (see, e.g. \cite{rodriguez2020computational}). Second, the STM has a broad ecosystem of tools and tutorials built around it, making our approach easier to replicate by other scholars.

We chose the number of topics in the model (parameter $k$) using a grid search over values of $k$ to find the model with the minimum perplexity and maximum log-likelihood, indicating the model with the best topic prediction probability on a held-out sample of words \cite{roberts2014structural}. We found the best $k$ to be 79. After the model was produced, we had two authors interpret the topics using the words grouped into each topic, and determine whether or not the topic was COVID-related or not. We calculated inter-annotator agreement for this task using Krippendorf's alpha \cite{krippendorffReliabilityContentAnalysis2004}, obtaining a value of 0.82. This is generally associated with substantial agreement, and high enough to use the construct in further analysis. The authors disagreed on 3 out of 79 topics, and settled on a final label after discussion. The top seven words for all topics in the model are given in Table~\ref{tab:STM3} in the Appendix.

\section{Results}

\begin{figure*}[t]
    \centering
    \includegraphics[width=\textwidth]{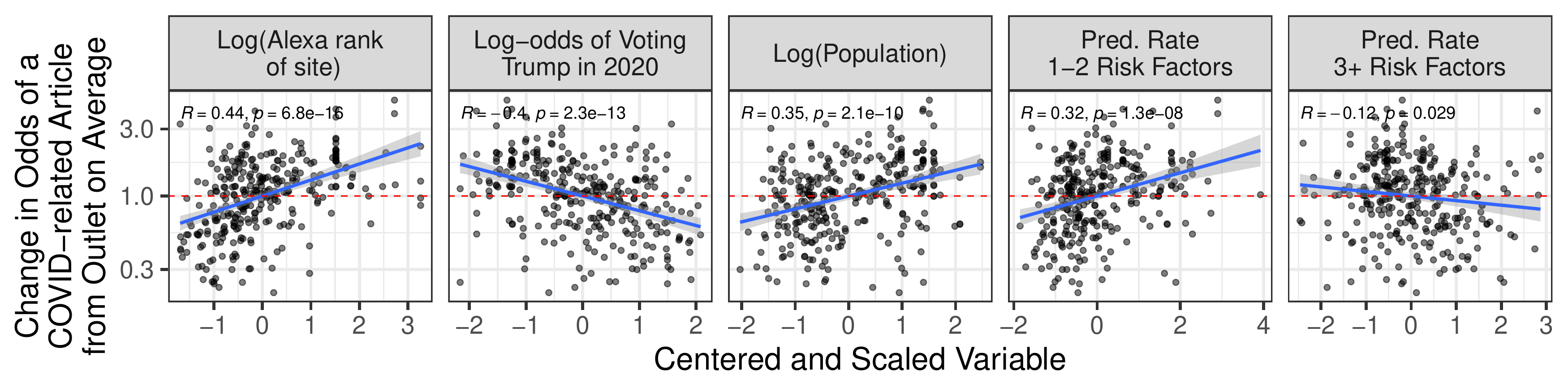}
    \caption{Each plot relates the change in the log-odds of a given outlet (each outlet is represented by a point) reporting on COVID (y-axis) to the centered and scaled distribution (x-axis) of a county- or outlet-level variable. The y-axis on these plots can be understood as the degree to which an outlet covered the pandemic, controlling for what we would expect given the variables in the regressions described above. The variables considered are defined in the grey heading for each subplot. From left to right, the variables considered are the (inverse) log of the news outlet website's ranking on Amazon Alexa's ranking of web domains, the log-odds of a person in the county in which the outlet is situated voting for Trump in 2020, the log of the county's population, and estimated percentage of the population with a particular number of COVID-related health risks. 
    }
    \label{fig:rq1c}
\end{figure*}
\subsection{RQ1: Associations with Overall Coverage}
\subsubsection{ Relationship of Coverage to COVID Cases and Deaths }
The proportion of local news coverage devoted to COVID was best explained by the number of 1) country-wide COVID deaths and cases, and 2) local case, but not death, rates of COVID. We also see that, controlling for these factors, local news outlets covered the pandemic significantly less as time wore on. Our results are based on a binomial regression model with fixed effects for each local news outlet. Our dependent variable is, for each local newspaper, the likelihood of a particular article written by that agency being about COVID in a given week.  Our independent variables are time, as characterized by the number of weeks since the start of 2020, and the logged rate per 1,000 people of observed new cases and deaths in the county, state, and country in which the local news outlet headquarters was situated.

The strongest predictor of COVID coverage by local news outlets was the (logged) number of total deaths across the U.S. in the week during which the article was written. As shown in Figure~\ref{fig:rq1a}, this relationship holds across a number of different model specifications; here we use the best fitting model according to model BIC to report results. According to this model, a one standard deviation increase in the number of country-wide deaths is associated with an 18\% (95\% clustered SE [17,19]) increase in the odds that a local news article at any outlet in our dataset will be about COVID. This relationship can be seen in the way that average coverage across all outlets tracks this country-level variable in Figure~\ref{fig:rq1b}.

In contrast, rates of death at local levels were not robustly associated with COVID coverage. Instead, robust effects at the local level were observed only in variations of case rates. State-level case rates were reliably predictive of COVID coverage across all models considered. A one standard deviation increase in cases in the state where a local outlet was embedded was associated with a 6.4\% [5.8,8.1] increase in coverage of the pandemic, relative to non-pandemic related content. State-level case rates predicted coverage better than county-level rates, but these two were correlated. In the alternative model presented in Figure~\ref{fig:rq1a}, for example, we find that county-level case rates have a significant effect when state-level cases are not included, although it is weaker than the state-level effect. 

\begin{figure*}[t]
    \centering
    \includegraphics[width=\textwidth]{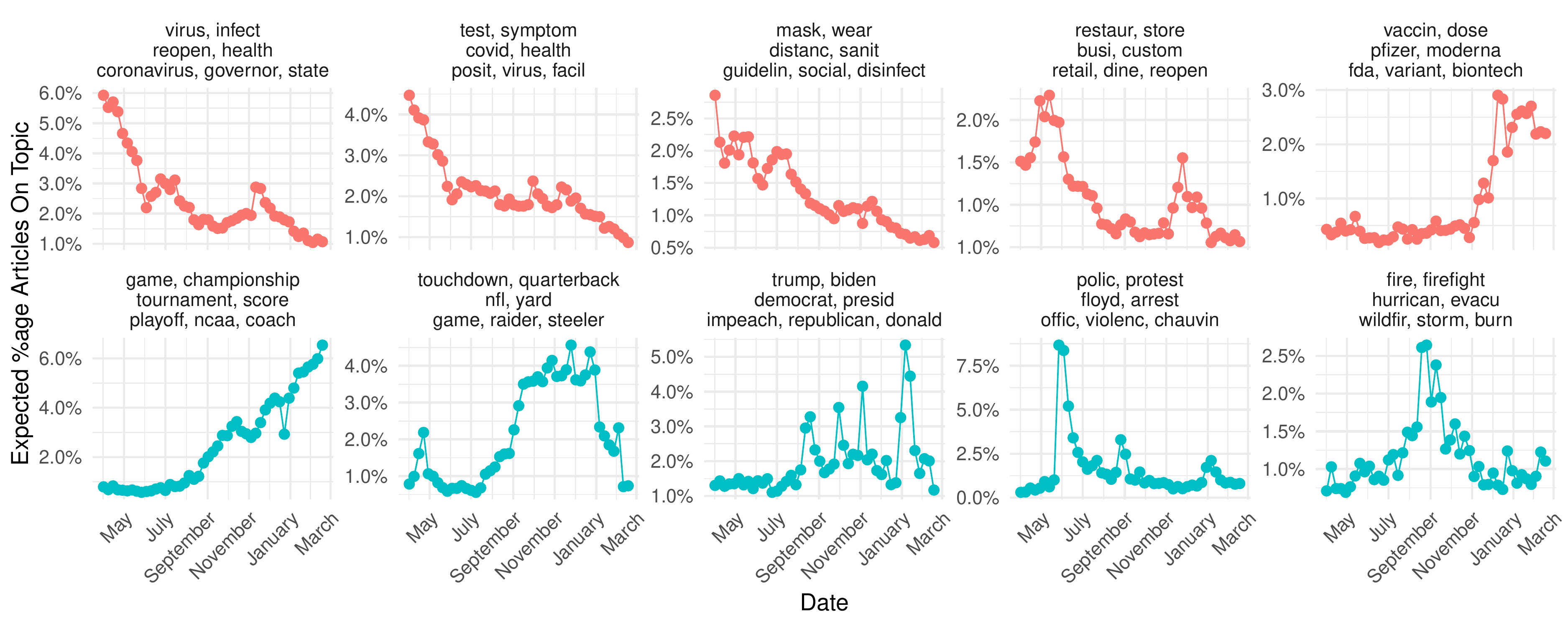}
    \caption{Each point on each subplot represents the expected percentage of articles in that week associated with five COVID-related topics (top row of subplots) and five non-COVID related topics (bottom row). Expected percentages are determined by a simple summation of the topic proportions of articles in each week as estimated by the STM. Topic labels are represented by the seven most probable terms for that topic. Topics displayed were selected to emphasize the diversity of temporal patterns across topics; results for all topics are provided in the associated GitHub repository.}
    \label{fig:rq2a}
\end{figure*}

Finally, we observe that both country-level case rates, as well as time, were also significantly associated with proportion of coverage. This means that a) an increase in country level case rates was associated with an increase in coverage, and that b) even though both case and death rates soared well after the pandemic began, local news outlets on average shifted attention away from the pandemic to other topics over time. Specifically, each additional week of time was associated with a 3.2\% [3.0,3.4] decrease in coverage, net of other factors.

These findings hold in a variety of alternative models in addition to those presented in Figure~\ref{fig:rq1a}. Specifically, in alternative models not presented here, we consider the possibility that the news articles that a given outlet produces in a given week are not impacted by case and death rates in the current week, but rather the previous one. To this end, we compare a model with no lagged variables (the model used here to report coefficients) with a model where both cases and deaths are lagged, where Cases are lagged but not Deaths, and where Deaths are lagged but not cases. Results for these models are provided in the online code and data release, but are not presented here due to space constraints. Two main findings arise from this comparison. First, the model with no lagged variables, i.e. the model whose results are described here, fit the data as well as or better than any model including lagged variables. Second, across all models, our main findings still hold---country-level deaths and cases are predictive of coverage, whereas at the local level, only state-level cases are predictive.  The main difference between the models is that cases become more important relative to deaths at the country level, which we do not focus on relative to the broader point of national-level focus in the main text.

\subsubsection{Variation in Overall Coverage Across Outlets} 

We find that outlets varied significantly in their coverage depending on the political leanings, population size, and health risk of their audience (as proxied by the county in which the outlet was situated), as well as the overall size of that outlet's audience.  Figure~\ref{fig:rq1c} presents bivariate relationships and correlation coefficients between these measures and the (centered) fixed-effect coefficient of each local news outlet from the regressions displayed in Figure~\ref{fig:rq1a}.  Centered fixed-effect coefficients are essentially a multiplier, such that a value of 3 can be read as ``this local news outlet was three times more likely to produce articles about COVID than the average outlet''.

Figure~\ref{fig:rq1c} shows that local news outlets in the most right-leaning counties in our dataset covered COVID around one tenth as much as the average outlet, and outlets in the most left-leaning counties around three times as much as the average outlet, controlling for local and national death and case rates, and time. Outlets with large audiences, both with respect to the population size of the county in which they were situated, and the outlet web page’s popularity, also covered COVID significantly more. 

Finally, areas that had a higher percentage of individuals with 1-2 COVID risk factors covered the pandemic significantly more, but areas with higher percentages of individuals with three or more COVID risk factors saw less coverage. Percentage of individuals with 1-2 risk factors was highly correlated with population size (.69), as certain factors were directly associated with increased population (e.g. overcrowding). This meant we could not reliably test the direct relationship between coverage and this level of risk, controlling for the other factors considered. However, correlation between percentage of individuals with three or more risk factors and the other explanatory variables considered was at most .10.  As such, we were able to test whether or not this effect could be explained by the other variables considered in Figure~\ref{fig:rq1c}. We find that this effect remains even after controlling for other county- and outlet-level factors. Specifically, we find that controlling for these other variables, a one standard deviation increase in the percentage of individuals with three or more COVID risk factors was associated with an 8.9\% [SE .03, p \textless .01] decrease in COVID coverage. This also held for the same analysis on fixed effects when only the limited filter was used (11.7\% [SE .04, p \textless .01]).

Results presented here use the fixed effects from the model used in the previous section to report coefficients. One possibility is that these fixed effects are, however, reliant on the model specification. Specifically, we consider the possibility that results may differ based on the keyword filter used. This does not appear to be the case. Specifically, fixed effects when using the limited filter instead of the full filter highly correlated (a Pearson Correlation of 0.81, p \textless .0001), and we therefore find that results are not sensitive to this change in modeling assumptions.

\subsection{RQ2:  Topic-specific coverage during the pandemic}

\begin{figure*}[t]
    \centering
    \includegraphics[width=\textwidth]{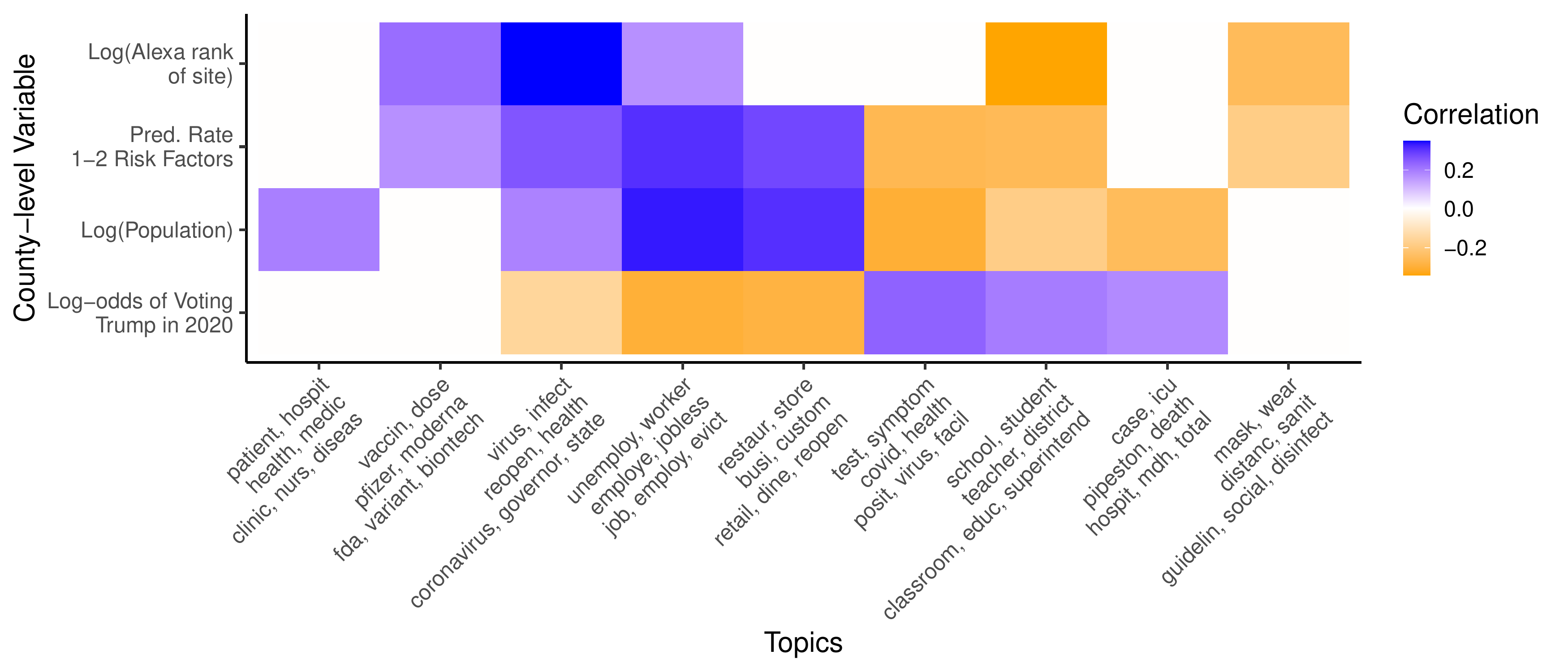}
    \caption{On the x-axis, the top 7 words for the nine COVID-related topics we identified. On the y-axis, the outlet and audience level variables we analyzed. Each cell of the plot is a relationship (using Pearson correlation) between the given outlet/audience level variable and the estimated percent of COVID-related coverage devoted to the given topic on the x-axis. Darker blue (orange) means a stronger positive (negative) correlation. Correlations that are not significant at p \textless .01 are displayed in white. We do not include the Rate of 3 or more risk factors, as there were no significant correlations with it.}
    \label{fig:rq2b}
\end{figure*}

Coverage of a variety of COVID-related topics peaked in the first few months of the pandemic. The top row of Figure~\ref{fig:rq2a} shows examples of the decline in coverage over time for three COVID-related topics, including topics related to masks and social distancing, state responses, and the economic impacts of the pandemic. However, this pattern was not monotonic. For example, discussions about cases and deaths from COVID showed an ebb and flow similar to the case and death rates themselves, and discussions about school reopenings naturally peaked at the beginning of the school year. Also, as one might expect, coverage of the vaccine began to rise rapidly in November 2020 as it became available in the United States. The bottom row of Figure~\ref{fig:rq2a} shows five non COVID-related topics that peaked after the initial wave of COVID in the United States, and represent both sudden and more gradual shifts in attention. More sudden displacements of COVID coverage came at critical times; for example, coverage of the George Floyd protests spiked in June, as cases were rising for the ``second wave'' of COVID in the U.S., and sports and election coverage similarly displaced COVID-related news during the early days of the deadliest peak of the virus.

However, we again find differences across outlets based on local politics, risk level, and outlet popularity. These results are summarized in Figure~\ref{fig:rq2b}, which displays correlations among the overall percentage of coverage an outlet devoted to a given COVID-related topic (x-axis) and four audience/outlet variables (y-axis). Note that topics displayed in Figure~\ref{fig:rq2b} represent \emph{all} COVID-related topics we identified. We find that articles about COVID written by outlets with more right-leaning, less populated audiences were more likely to be related to testing, school closures, and/or case and death counts. In contrast, outlets with larger, more left-leaning audiences focused more heavily on state-level politics and economic impacts in the form of unemployment, housing, and restaurant re-opening. 

We also see that outlets with more popular websites had higher rates of vaccine-related coverage, and less coverage about masks and social distancing. Finally, the percentage of an outlet's audience with three or more risk factors was not significantly associated with coverage of any particular topic. Percentage of audience with 1-2 risk factors was associated with a variety of topics in ways that would be expected given its association with county population size. 

\section{Limitations}

While we have worked to ensure that our analysis is robust to a number of potential issues, our work nonetheless contains several important limitations. First and foremost are potential limitations with our sampling of local news sites. Our sample does cover a broad geographic area and a diverse array of communities within the U.S., and is, again, the largest sample we are aware of for local news outlets. However, it relies on a specific definition of local news that relies on subjective decisions about whether or not a site is "local." For example, we consider the New York Times a national outlet, because of the well-known national readership of the paper, but not the Buffalo News. Our sampling also relies on local news outlets listed by a well-established, but nonetheless singular directory of sites. Finally, we analyze only local news outlets with RSS feeds, not those with paywall mechanisms, which may bias our sample towards less well-established outlets.

Second, our analysis also cannot establish any causal relationships, instead allowing us only to understand associations among impacts of the virus, local news coverage, and the public’s views. This is relevant, for example, in the proxy variables we select; we chose to use only variables we could motivate theoretically, and not, for example, potentially important causal factors such as the location of an outlet in different parts of the country. Finally, given space constraints, we are only able to focus on a descriptive analysis of local news outlet content from our dataset, and thus do not present a picture of how our data compares to other, more limited local news datasets, nor how local news coverage compares to coverage from national outlets, or to discussion on social media. Both of these analyses are fruitful avenues of future work; as our data is available, we hope that such analyses can be carried out by ourselves and others.

\section{Conclusion}

Local news plays a fundamental role in the provision of information to Americans. However, our findings show that what local news agencies covered was best explained not by local conditions, but rather by national trends in the number of people dying from COVID. This finding reflects a continuing trend of the nationalization of news coverage, a trend that is decades in the making, but has gotten more pronounced in recent years \cite{schudson1995power,hayesLocalNewsGoes2015}. Assessing local news capacity and the relative “information needs of communities” remain crucial and growing areas of research, and the current study contributes to that literature \cite{napoli2017local}. 

We also find that collectively, local news outlets failed to return a focus to COVID as a primary theme of coverage as the degree of tragedy in November of 2020 rose and eventually greatly surpassed the case and death rates of the Spring of 2020. This coverage fatigue, while not a new phenomenon, shows yet another vector along which Americans were deprived of coverage of the pandemic that may have helped to inform their behavior. Further, we see that the coverage of COVID that local news outlets did provide was not evenly distributed, in two respects. First, the degree to which outlets covered COVID differed based on local politics, outlet popularity, and audience risk level. Perhaps most notably, we found that outlets with larger percentages of the most at-risk individuals provided significantly less coverage.

Second, even with similar rates of overall coverage, outlets differed in what themes they focused on. One surprising observation in this respect is that outlets with more left-leaning audiences focused more on the political and economic impacts of the pandemic, whereas those with right-leaning audiences focused more heavily on school closures and the more health-related themes of testing and case counts. These relationships can be explained in part by the connection between politics and population size; economic impacts in particular were likely more acute for larger cities, and as such drew coverage away from more universal themes of school closing and health-related issues in these areas. Importantly, however, this shift in focus was often offset by a larger amount of COVID-related coverage.

These observations, overall, suggest that local news outlets may not have provided the degree of coverage that COVID warranted given local conditions, and that coverage that did occur was tailored at times to fit the desires, rather than the needs, of their audience. These claims should be taken in concert with the limitations of this study discussed above. Nonetheless, our work shows the potential utility of engaging in real-time monitoring of information being produced at the most local levels of individuals’ information environments. 

\begin{small}
\bibliography{scibib}
\end{small}

\appendix

\section{Appendix}
Table~\ref{tab:STM3} presents the top seven words for each of the 79 topics generated by our structural topic modeling analysis presented in the main text. 

\begin{table*}[h!]
\scriptsize
\begin{tabular}{l|p{7.cm}|l|p{7.cm}}
\toprule
\small 
\#   &  Highest Prob Words & \#   &  Highest Prob Words \\
\toprule
 1 & suppli, equip, shortag, manufactur, associ, ventil, product & 2 & senat, republican, democrat, lawmak, legisl, gop, sen \\
 3 & north, dakota, carolina, south, fargo, burgum, fork & 4 & tax, fund, budget, loan, revenu, billion, million \\
 5 & weather, snow, wind, storm, winter, rain, temperatur & 6 & peoria, columbus, anderson, adam, zoo, lion, kona \\
 7 & feel, thing, lot, need, someth, work, peopl & 8 & minist, iran, russia, navalni, govern, russian, britain \\
 9 & polici, state, enforc, law, public, requir, violat & 10 & may, april, june, march, date, juli, avg  \\
11 & paul, marshal, pardon, stone, sacramento, twin, wall  & 12 & ballot, vote, elect, voter, absente, democrat, elector  \\
13 & light, earth, sun, moon, sky, blue, planet  & 14 & virus, infect, reopen, health, coronavirus, governor, state  \\
15 & oct, sept, aug, nov, dec, star, jan  & 16 & moor, kennedi, lgbtq, transgend, duluth, gay, mcdonald  \\
17 & mask, wear, distanc, sanit, guidelin, social, disinfect  & 18 & water, oil, emiss, energi, gas, environment, electr  \\
19 & coach, athlet, footbal, sport, season, player, school  & 20 & student, graduat, colleg, univers, campus, scholarship, semest  \\
21 & properti, project, construct, estat, acr, build, street  & 22 & day, weekend, hour, memori, morn, everi, arriv  \\
23 & funer, grandchildren, natchez, gravesid, cemeteri, church, vicksburg  & 24 & polic, vehicl, sheriff, man, crash, investig, shoot  \\
25 & island, port, beach, boat, insle, clallam, harbor  & 26 & communiti, donat, program, food, support, servic, nonprofit  \\
27 & triblivecom, wpial, township, allegheni, pittsburgh, tribun  & 28 & book, publish, read, edit, reader, editor, print  \\
29 & polit, govern, democraci, politician, peopl, truth, opinion  & 30 & case, icu, pipeston, death, hospit, mdh, total  \\
31 & counti, sheriff, commission, depart, yolo, resid, offic  & 32 & plant, farm, cbd, farmer, tree, agricultur, garden  \\
33 & app, user, amazon, technolog, googl, tiktok, internet  & 34 & council, board, citi, town, mayor, meet, approv  \\
35 & arrest, jail, prison, inmat, feloni, assault, sentenc  & 36 & meal, food, chees, bake, chicken, sauc, wine  \\
37 & event, festiv, parad, cancel, club, ticket, celebr  & 38 & touchdown, quarterback, nfl, yard, game, raider, steeler  \\
39 & test, symptom, covid, health, posit, virus, facil  & 40 & password, subscrib, subscript, premium, log, regist, inquir  \\
41 & compani, industri, market, busi, invest, ceo, investor  & 42 & got, guy, mayb, walk, kid, fun, car  \\
43 & inning, dodger, basebal, mlb, game, sox, player  & 44 & vaccin, dose, pfizer, moderna, fda, variant, biontech  \\
45 & plan, phase, walz, schedul, announc, decis, minnesotan  & 46 & payment, applic, postal, mail, census, dejoy, servic  \\
47 & new, york, cuomo, mexico, jersey, schenectadi, rhode  & 48 & race, nascar, championship, runner, olymp, hors, harvick  \\
49 & game, championship, tournament, score, playoff, ncaa, coach & 50 & fire, firefight, hurrican, evacu, wildfir, storm, burn  \\
51 & black, racism, racial, statu, confeder, african, racist  & 52 & dog, anim, pet, cat, shelter, havr, homeless  \\
53 & church, christma, holiday, god, worship, pastor, jesus  & 54 & militari, veteran, war, armi, navi, soldier, troop  \\
55 & baker, boston, pritzker, vineyard, poli, denver, springfield  & 56 & school, student, teacher, district, classroom, educ, superintend  \\
57 & children, women, child, parent, girl, kid, age & 58 & unemploy, worker, employe, jobless, job, employ, evict  \\
59 & restaur, store, busi, custom, retail, dine, reopen  & 60 & una, para, más, pero, sus, también, dijo  \\
61 & argus, dispatch, whbf, quad, wvik, fax, kwqc    & 62 & china, airlin, flight, chines, airport, spacex, korea  \\
63 & film, music, movi, song, album, theater, actor    & 64 & court, attorney, lawsuit, suprem, justic, judg, alleg  \\
65 & rate, percent, economi, million, increas, price, averag   & 66 & vega, casino, hotel, clark, tribe, sisolak, carson  \\
67 & san, newsom, santa, francisco, diego, immigr, jose   & 68 & love, famili, father, mother, son, daughter, dad  \\
69 & librari, https, onlin, zoom, registr, inform, virtual    & 70 & art, artist, museum, galleri, paint, mural, exhibit  \\
71 & polic, protest, floyd, arrest, offic, violenc, chauvin  & 72 & trump, biden, democrat, presid, impeach, republican, donald  \\
73 & nba, game, player, laker, nhl, playoff, season   & 74 & patient, hospit, health, medic, clinic, nurs, diseas  \\
75 & park, ski, trail, aspen, boulder, bike, steamboat    & 76 & sexual, dear, abus, mental, abbi, suicid, violenc  \\
77 & award, smith, honor, sidney, hall, fame, winner   & 78 & fish, lake, wildlif, angler, speci, bird, hunt  \\
79 & year, last, first, month, run, sinc, ago  \\
\end{tabular}	
\caption{The 79 topics generated by the structural topic model used in the paper. Topic numbers are presented alongside the seven words that were most probable in the topic's posterior distribution. Note, words are stemmed.}
\label{tab:STM3}
\end{table*}

\onecolumn
\section{Supplement}

In this supplement, we provide additional information on alternative model specifications for our analysis of COVID coverage rates, and additional details on the output of our topic modeling.

\begin{figure}[t]
    \centering
    \includegraphics[width=\textwidth]{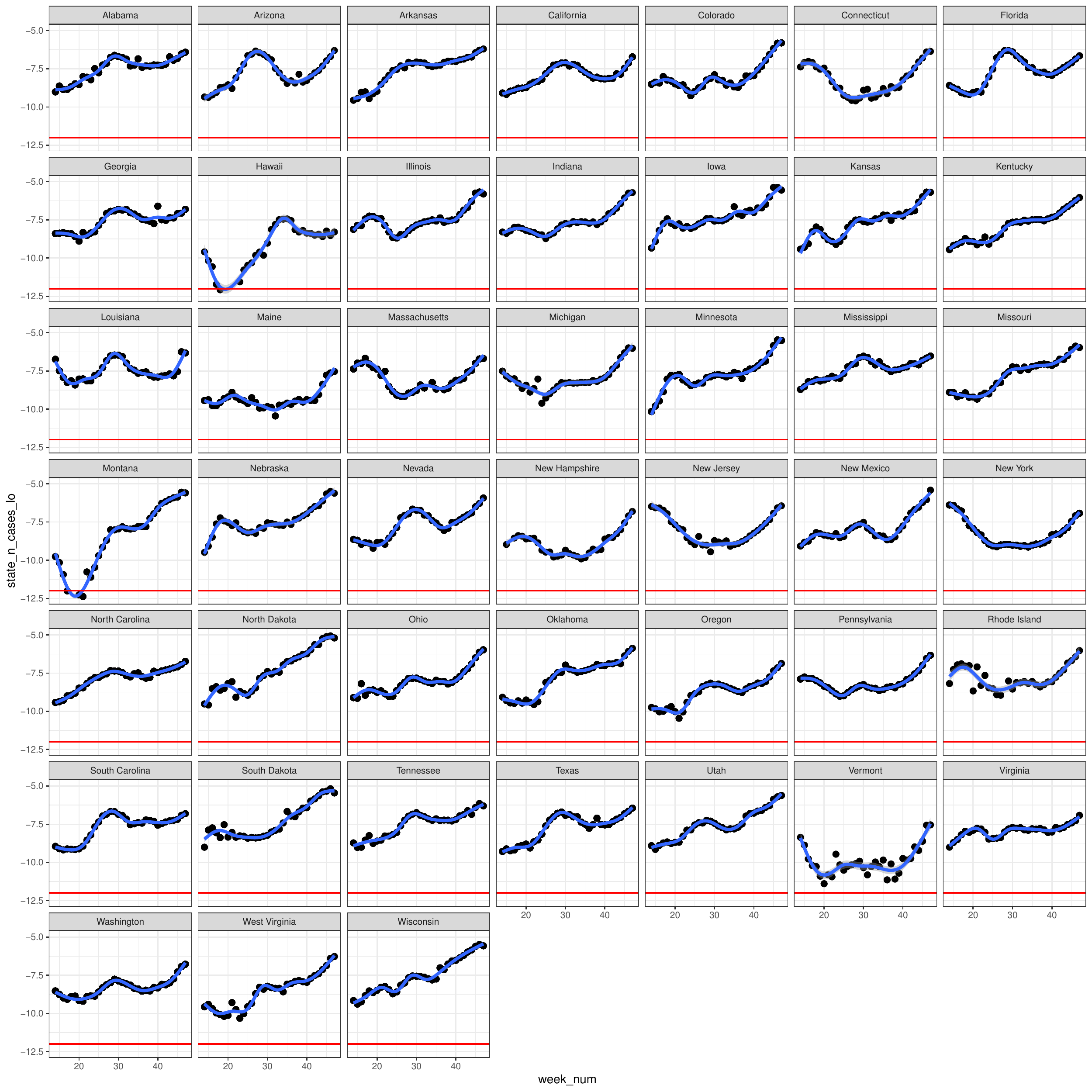}
    \caption{Weekly case rates per 1,000 people at the state level}
    \label{fig:state_cases}
\end{figure}

\begin{figure}[t]
    \centering
    \includegraphics[width=\textwidth]{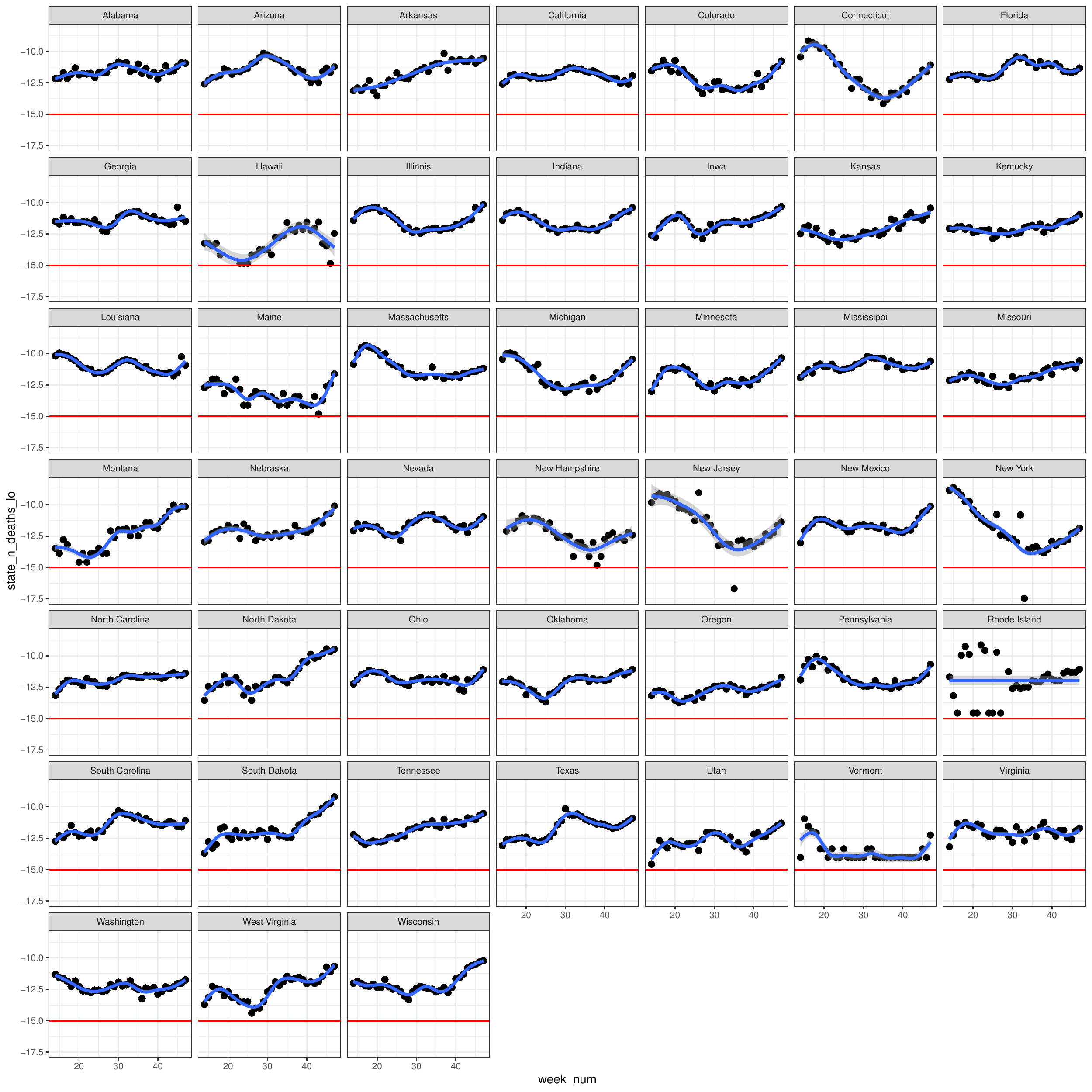}
    \caption{Weekly death rates per 1000 people at the state level}
    \label{fig:state_deaths}
\end{figure}

\begin{figure}[t]
    \centering
    \includegraphics[width=\textwidth]{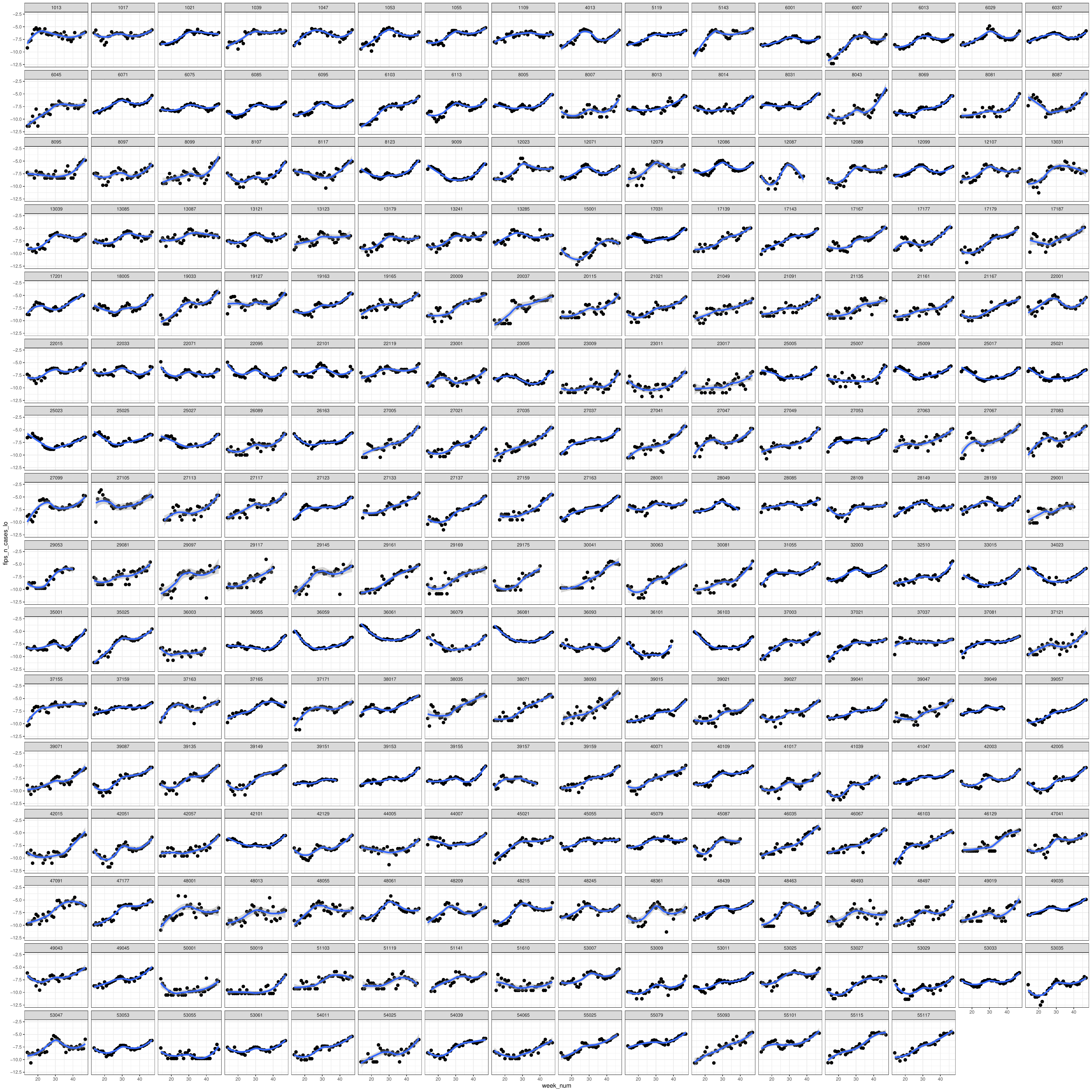}
    \caption{Weekly case rates per 1,000 people at the state level}
    \label{fig:counties_cases}
\end{figure}

\begin{figure}[t]
    \centering
    \includegraphics[width=\textwidth]{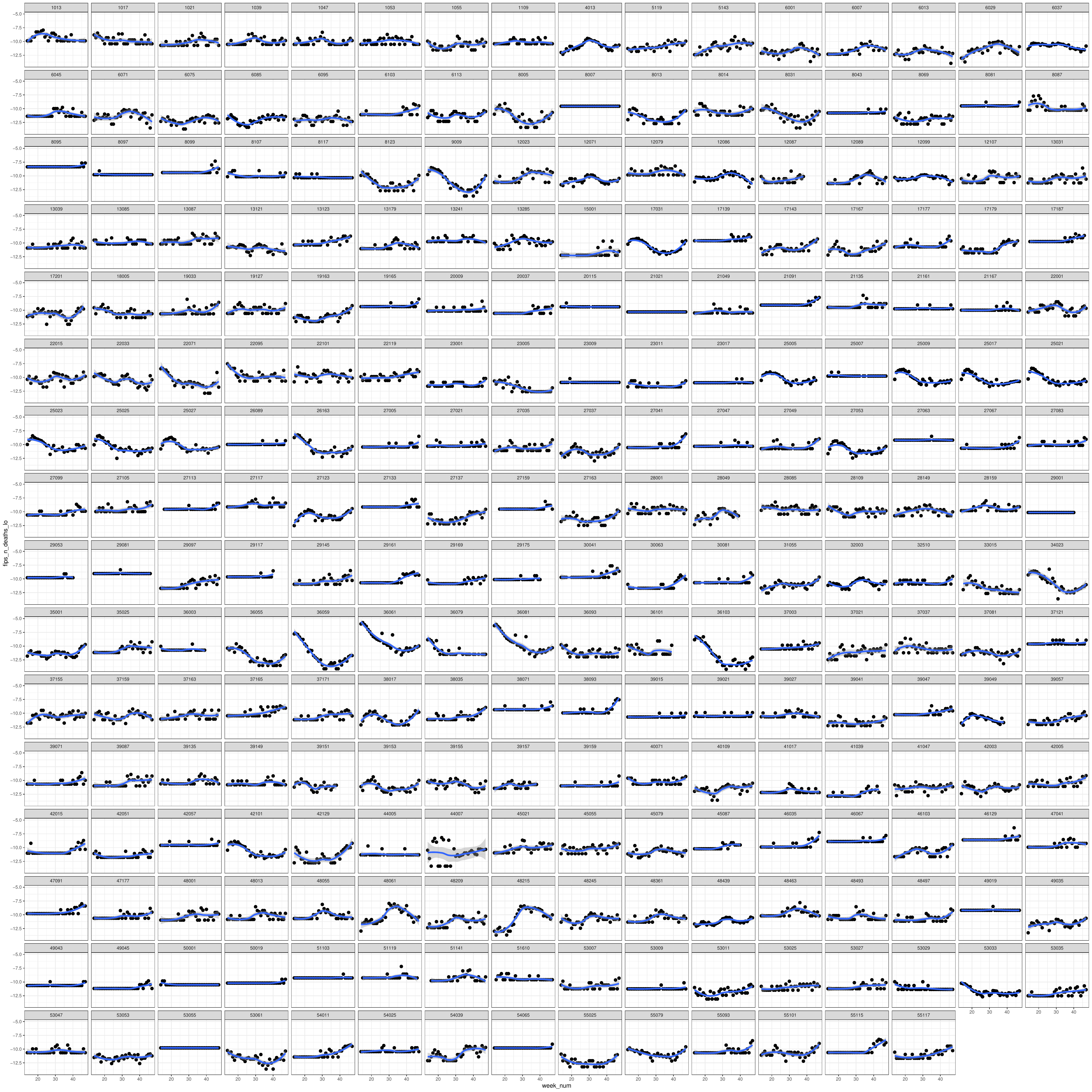}
    \caption{Weekly death rates per 1,000 people at the state level}
    \label{fig:counties_deaths}
\end{figure}
\section{Alternative Models for COVID Coverage Rates}\label{sec:models}

We provide details of several alternative models that show the robustness of our main findings to several potential differences in specification. 

First, Table~\ref{tab:mods} presents the four models described in the main text. Model 1 is the model used in the main text to report coefficients. Model 2 is the same model but with the limited COVID keyword filter, as opposed to the full filter. Models 4 and 5 were run without state (Model 4) or County (Model 5) variables to address the potential that correlations between state and county-level case and death rates (particularly for large counties) might hide effects at the local level. All four of these models, as shown in the main text, produce the same substantive conclusions; results are presented here in an alternative format and with additional statistics on model fit. The one model not presented in the main text in Table~\ref{tab:mods} is Model 3. This model shows that our results are robust to the removal of a cluster of 20 news sites in Massachusetts all run by the same parent company (WickedLocal). While these sites produce different content, it is possible that they could skew our results. However, Table~\ref{tab:mods} shows that this is not the case.

Second, one might imagine that the news articles an outlet produces in a given week are not impacted by case and death rates in the current week, but rather the previous one. To this end, in Table~\ref{tab:lag}, we compare a model with no lagged variables (Model 1, the same used in the main text to report coefficients) with a model where both cases and deaths are lagged (Model 2), where Cases are lagged but not Deaths (Model 3), and where Deaths are lagged but not cases (Model 4). Two main findings arise from this comparison. First, the model with no lagged variables, i.e. the model used in the main text, is the most predictive model according to Squared Correlation, whereas the model with lagged death rates is preferred according to BIC. Second, however, we find that amongst these two most predictive models, our main findings still hold---country-level deaths and cases are predictive of coverage, whereas at the local level, only state-level cases are predictive.  The main difference between the models is that cases become more important relative to deaths at the country level, which we do not focus on relative to the broader point of national-level focus in the main text.

\begin{figure}
    \centering
    \includegraphics[width=.8\textwidth]{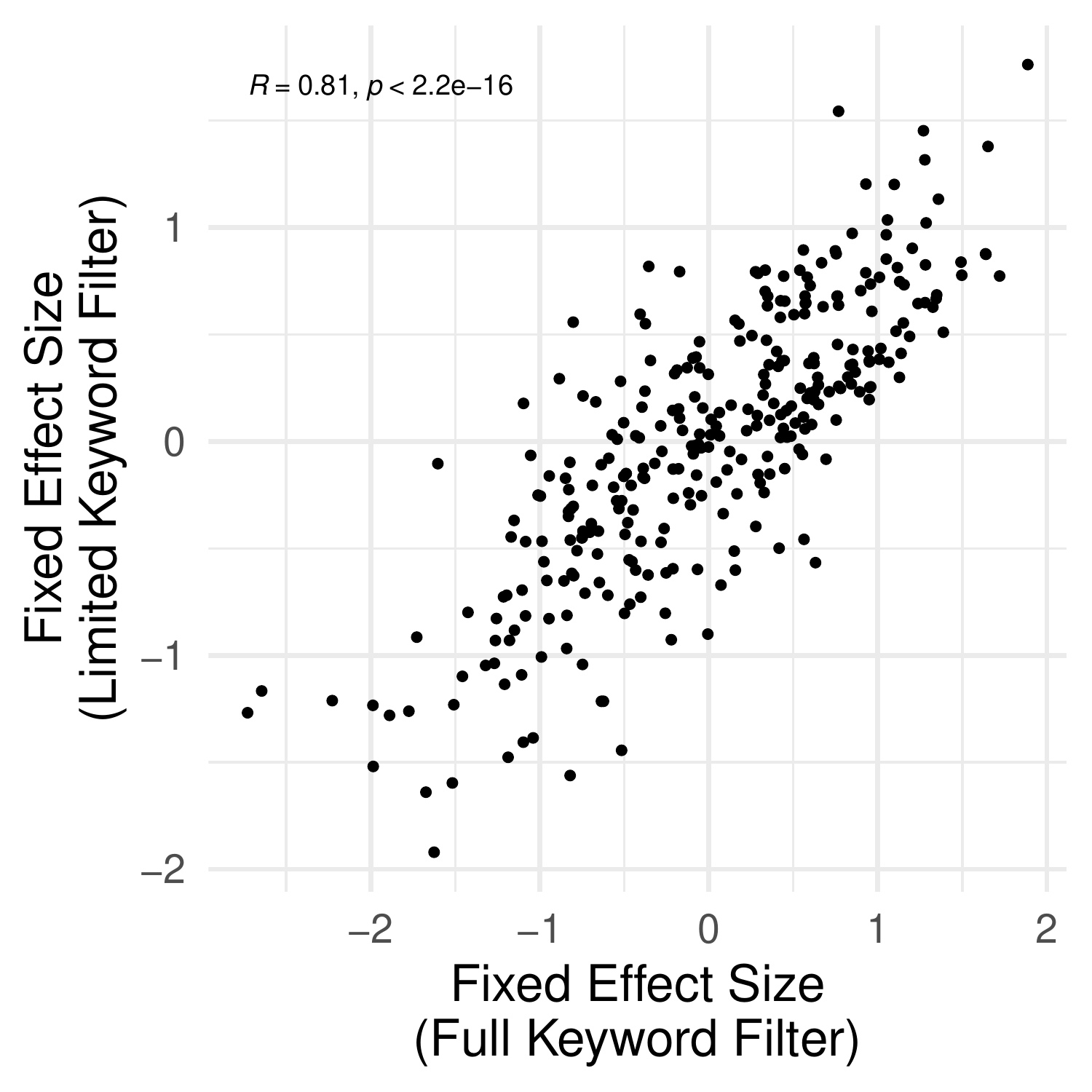}
    \caption{Correlation between fixed effects with the Limited vs Full filter. Because these are strongly correlated, we provide only an analysis of results with the full filter.}
    \label{fig:lim_vs_full_fe}
\end{figure}

Finally, our analysis in the main text also uses the fixed effects from Model 1 (the same model in both Table~\ref{tab:mods} and Table~\ref{tab:lag}) to analyze differences at the outlet level. One possibility is that these fixed effects are reliant on the model specification. Specifically, results may differ based on the keyword filter used. In Figure~\ref{fig:lim_vs_full_fe}, we therefore plot the fixed effects of all outlets, comparing results when the full versus the limited keyword filter is used. The figure shows that these two quantities are highly correlated (a Pearson Correlation of 0.81, p \textless .0001), and thus that these results are not sensitive to this change in modeling assumptions.

\begin{table}
\centering
\begin{tabular}{lccccc}
\tabularnewline\midrule\midrule
Keyword Filter:&Full& Limited& Full& Full& Full\\
Addt'l Points:& & & No MA & No State & No County\\
Model:&(1) & (2) & (3) & (4) & (5)\\
\midrule \emph{Variables}&   &   &   &   &  \\
Weeks Since 1/1/20&-0.0320$^{***}$ & -0.0365$^{***}$ & -0.0331$^{***}$ & -0.0309$^{***}$ & -0.0319$^{***}$\\
  &(0.0016) & (0.0018) & (0.0017) & (0.0015) & (0.0016)\\
 N. Cases Country-level &0.0343$^{***}$ & 0.0506$^{***}$ & 0.0358$^{***}$ & 0.0483$^{***}$ & 0.0355$^{***}$\\
  &(0.0100) & (0.0114) & (0.0102) & (0.0099) & (0.0100)\\
 N. Cases County-level &0.0233 & -0.0112 & 0.0253$^{*}$ & 0.0571$^{***}$ &   \\
  &(0.0149) & (0.0179) & (0.0151) & (0.0133) &   \\
 N. Cases State-level &0.0644$^{***}$ & 0.1099$^{***}$ & 0.0717$^{***}$ &    & 0.0789$^{***}$\\
  &(0.0167) & (0.0170) & (0.0171) &    & (0.0161)\\
 N. Deaths Country-level &0.1801$^{***}$ & 0.1811$^{***}$ & 0.1848$^{***}$ & 0.1837$^{***}$ & 0.1790$^{***}$\\
  &(0.0093) & (0.0114) & (0.0093) & (0.0094) & (0.0094)\\
 N. Deaths State-level &-0.0077 & -0.0336$^{**}$ & -0.0043 &    & -0.0111\\
  &(0.0138) & (0.0151) & (0.0148) &    & (0.0128)\\
 N. Deaths County-level &-0.0073 & -0.0056 & -0.0051 & -0.0023 &   \\
  &(0.0090) & (0.0109) & (0.0086) & (0.0083) &   \\
\midrule \emph{Fixed-effects}&   &   &   &   &  \\
sourcedomain\_id & Yes & Yes & Yes & Yes & Yes\\
\midrule \emph{Fit statistics}&  & & & & \\
Observations & 10,327&10,327&9,309&10,327&10,327\\
Squared Correlation & 0.65524&0.68429&0.64845&0.65453&0.65488\\
Pseudo R$^2$ & -3.0756&-3.6748&-3.1199&-3.0829&-3.0768\\
BIC & 61,258.1&55,359.7&55,666.1&61,344.0&61,255.7\\
\midrule\midrule\multicolumn{6}{l}{\emph{One-way (sourcedomain\_id) standard-errors in parentheses}}\\
\multicolumn{6}{l}{\emph{Signif. Codes: ***: 0.01, **: 0.05, *: 0.1}}\\
\end{tabular}
\caption{Models 1,2,4, and 5 are as presented in the main text. Model 4 is the same as Model 1, but removes a cluster of news outlets in Massachusetts all run by the same agency}
\label{tab:mods}
\end{table}

\begin{table}
\centering
\begin{tabular}{lcccc}
\small
\tabularnewline\midrule\midrule
Dependent Variable:&\multicolumn{4}{c}{Percentage of COVID-related Articles}\\
Model:&(1) & (2) & (3) & (4)\\
Weeks Since 1/1/20&-0.0320$^{***}$ & -0.0399$^{***}$ & -0.0275$^{***}$ & -0.0427$^{***}$\\
  &(0.0016) & (0.0016) & (0.0016) & (0.0016)\\
 N. Cases Country-level &0.0343$^{***}$ &    &    & 0.1255$^{***}$\\
  &(0.0100) &    &    & (0.0104)\\
 N. Cases County-level &0.0233 &    &    & 0.0211\\
  &(0.0149) &    &    & (0.0164)\\
 N. Cases State-level &0.0644$^{***}$ &    &    & 0.0764$^{***}$\\
  &(0.0167) &    &    & (0.0185)\\
 N. Deaths Country-level &0.1801$^{***}$ &    & 0.1915$^{***}$ &   \\
  &(0.0093) &    & (0.0100) &   \\
 N. Deaths State-level &-0.0077 &    & 0.0028 &   \\
  &(0.0138) &    & (0.0135) &   \\
 N. Deaths County-level &-0.0073 &    & -0.0086 &   \\
  &(0.0090) &    & (0.0091) &   \\
 N. Cases Country-level ({\bf Lag 1}) &   & 0.1032$^{***}$ & 0.0094 &   \\
  &   & (0.0093) & (0.0095) &   \\
N. Cases County-level ({\bf Lag 1}) &   & 0.0277$^{**}$ & 0.0284$^{**}$ &   \\
  &   & (0.0136) & (0.0116) &   \\
N. Cases State-level ({\bf Lag 1}) &   & 0.0519$^{***}$ & 0.0375$^{***}$ &   \\
  &   & (0.0161) & (0.0134) &   \\
N. Deaths Country-level ({\bf Lag 1}) &   & 0.1116$^{***}$ &    & 0.1293$^{***}$\\
  &   & (0.0077) &    & (0.0075)\\
N. Deaths State-level ({\bf Lag 1}) &   & -0.0102 &    & -0.0161\\
  &   & (0.0140) &    & (0.0127)\\
N. Deaths County-level ({\bf Lag 1})&   & -0.0063 &    & -0.0026\\
  &   & (0.0096) &    & (0.0091)\\
\midrule \emph{Fixed-effects}&   &   &   &  \\
News Outlet & Yes & Yes & Yes & Yes\\
\midrule \emph{Fit statistics}&  & & & \\
Observations & 10,327&10,327&10,327&10,327\\
Squared Correlation & 0.65524&0.63793&0.65403&0.64340\\
BIC & 61,258.1&62,870.4&61,459.1&62,173.1\\
\multicolumn{5}{l}{\emph{Signif. Codes: ***: 0.01, **: 0.05, *: 0.1}}\\
\end{tabular}
\caption{Comparison with Lagged Variables. All Variables are rates per 1,000 people, and are logged, centered, and scaled by one standard-deviation. Standard Errors are provided, clustered by news outlet.}
\label{tab:lag}
\end{table}

%
%
%

\end{document}